\documentclass[11pt]{amsart}
\usepackage{geometry}                
\usepackage{graphicx}
\usepackage{amssymb}
\usepackage{epstopdf}

\usepackage{amsaddr}

\title{Folded Goursat surface and banana-shaped seedpod}
\author{Etienne Couturier}
\address{Laboratoire MSC Universite Paris-Diderot}
\email{etienne.couturier@univ-paris-diderot.fr}

\begin{document}
\maketitle
\begin{abstract}
Thin vegetal shells have recently been a \text{sign}ificant source of inspiration for the de\text{sign} of smart materials and soft actuators. Herein is presented a novel analytical family of isometric deformations with folds, inspired by a banana-shaped seedpod, which converts a vertical closing into either an horizontal closing or an opening depending on the location of the fold. Similarly to the seedpod,  optimum shapes for opening ease are the most elongated ones.
\end{abstract}
\section{Introduction}
\begin{figure}[h!]
\centering\includegraphics[width=5in]{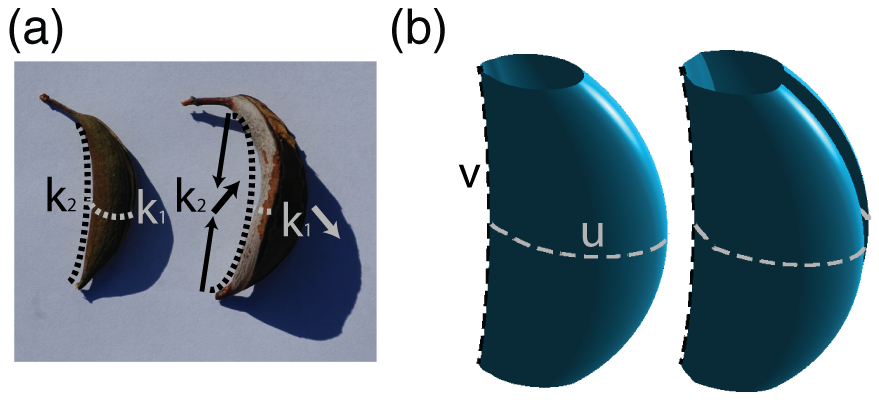} 
\caption{(a) (Left) A sealed and turgid seedpod of $\mathit{Acaccia\ caven}$. (Right) An open and desiccated seedpod. $k_1$, $k_2$ stand for the meridional and the longitudinal   curvatures at the saddle point. During desiccation the shortening of some external fibers at the back could increase $k_2$ and by consequence as the product $k_1 k_2$ is conserved for an isometry $k_1$ should decrease; it would thus actuate the opening of the seedpod. (b) A folded Goursat surface behaving similarly to the seedpod. ($R_1=0.58$, $\alpha_1=\alpha_2= 0.7752$, $b=0.5$, $c=10$, $d=30$, $\epsilon=\text{sign}(u_{jun,2}-u)$, $v_0=0.94$). (Left) Rest state $h=0$. (Right) Deformed state $h=0.05$. $u$ (gray dashed line) and $v$ (black dashed line) are the two coordinates lines of the system presented in Section 3a crossing at the saddle point.}
\label{Figure_1}
\end{figure}

Thin vegetal shells and rods have recently been a \text{sign}ificant source of inspiration for the de\text{sign} of smart materials and soft actuators: pinecone for bending actuator \cite{1}, orchid seedpod \cite{2} and the seed of $\mathit{Erodium}$ for twisting actuator  \cite{3}. Recent progress in chemistry and in synthesis of fibrous material have made possible to reproduce their behavior in biomimetic devices \cite{4}. Theoretical progress involving differential geometry has procured a deeper insight into the principles of these structures \cite{3},\cite{5},\cite{6}. \\
Herein is  described another kind of smart shells inspired by the banana-shaped seedpod of $\mathit{Accacia\ caven}$ from southern Chile. While dessicating, the longitudinal curvature at the saddle point of the banana increases while its meridional curvature decreases which triggers the opening of the banana shaped seedpod and enables seed-dispersal  (Figure \ref{Figure_1}a-b). It is a classical result that the most energetically economical modes of deformation for thin shells are the isometric ones when they are possible \cite{7}. The local scenario is compatible with isometry as the product of the principal curvatures (Gaussian curvature) at the saddle point could be kept constant if both curvatures vary inversely. Proving this local scenario can be extended into a global solution for the whole surface is in general a complex problem of PDE \cite{8}. Another possible approach is to construct an analytical solution. \\
While looking for such a solution, we ended up with a novel family of $C^0$ isometric deformation surfaces generalizing the classical Goursat family \cite{9} by naturally adding folds. Our family of surfaces includes banana-shaped surfaces which behave similarly to the seedpod at the saddle point -- increase in longitudinal curvature induces meridional opening, while the fold antagonistically tends to close the shell; depending on the fold location, either the opening component or the closing component will dominate. In this article we provide two examples: a $C^\infty$ family of shells for which an increase of the longitudinal curvature at the saddle point does not trigger the opening but triggers the closing of the aperture instead, and a  biomimetic $C^1$ family of folded shells for which an increase of the longitudinal curvature at the saddle point does indeed trigger the opening. Shape optimization in the latter family, easily carried on thanks to the analytic formulation, shed a new light on the elongated seedpod de\text{sign} which minimizes  the cost both in energy and in longitudinal deformation for the opening. \\

\section{Goursat surface with a fold and mechanical energy}
\subsection{Goursat surface}
In 1891, Goursat discovered the widest family of surfaces which can be isometrically deformed two orthogonal systems of parallel planes being preserved ((8) to (12) of \cite{9}); ten years later, Raffy \cite{10} even proved this family could not be enlarged requiring only one system of parallel planes to be preserved. Let $u_{min}<u_{max}$, $v_{min}<v_{max}$, $I=[u_{min},u_{max}]$, $J=[v_{min},v_{max}]$, $h>0$, $U_1$, $U_2$, $U_3$ three real functions on $I$ and $V_1$, $V_2$  two real functions on $J$.  The Goursat family \cite{9} can be written:
\begin{align}\label{1.1.1}
\begin{split}
 \forall h\geqslant0,\forall u\in I,\forall v&\in \Big(J\cap (V_2^{\prime2}-h V_1^{\prime2})^{-1}(]0,+\infty[) \Big),\\
S(u,v,h)&=V_1(v) f(u,h)  e_r +\int_{t=0}^{t=u} U_3(t) \Gamma(t,h) dt+   \Big( \int_0^v  \sqrt{V_2^{\prime2}-h V_1^{\prime2}} dv\Big)e_z
\end{split}
\end{align}

\noindent where $f$, $\theta$, $e_r$, $e_\theta$, $e_z$ and $\Gamma$ read\footnote{$\mathbf{Notations}$: For a function f, $\frac{\partial f}{\partial u}$ is notated $f_u$}:
\begin{align}\label{1.1.2}
\begin{split}
 \forall h \geqslant 0,\forall u\in I,\\
 f(u,h)&=\sqrt{U_1(u)^2+U_2(u)^2+h}\\
\theta(u,h)&=\int_0^{u} \frac{\sqrt{(U_1 U_2'-U_2 U_1')^2+h(U_1^{\prime2}+U_2^{\prime2})}}{f^2} du\\
e_r(u,h)&=(\cos(\theta(u,h)),\sin(\theta(u,h)),0)\\
e_\theta(u,h)&=(-\sin(\theta(u,h)),\cos(\theta(u,h)),0)\\
e_z&=(0,0,1)\\
\Gamma(u,h)&=\frac{\partial (f e_r)}{\partial u}(u,h)=f_u(u,h) e_r(u,h)+\theta_u(u,h) f(u,h) e_\theta(u,h)
\end{split}
\end{align}
$u$, $v$ are the coordinates (see Figure\ref{Figure_1} b) and $h$ is the parameter of deformation. An example of Goursat surface is represented on the top panel of Figure \ref{Figure_2} b. The set where $v$ can be chosen is smaller than $J$ and depends on $h$ because if $h$ is superior to $ h_{\max}=\frac{V_2'(v)^2}{V_1'(v)^2}$, $V_2'(v_0)^2-h V_1'(v_0)^2$ is negative and thus the surface partly imaginary.

\subsection{Isometric Goursat deformation}
\noindent As the article of Goursat is old and french-written, the proof that the deformations are isometric is recapitulated. For $h>0$, for $u\in I$ and for $v\in \Big(J\cap (V_2^{\prime2}-h V_1^{\prime2})^{-1}(]0,+\infty[) \Big)$, the first fundamental form reads: 
\begin{align}\label{1.2.1}
\begin{split}
E&=S_u^2\\
F&=S_u\cdot S_v\\
G&=S_v^2
\end{split}
\end{align}

\noindent where $S_u$ and $S_v$ are
\begin{align}\label{1.2.2}
\begin{split}
S_u&=(V_1+U_3) ( f_u e_r+f\theta_u e_{\theta})\\
S_v&= f V_1' e_r+  \sqrt{V_2^{\prime2}-h V_1^{\prime2}} e_z
\end{split}
\end{align}
with:
\begin{align}\label{1.2.3}
\begin{split}
f_u&=\frac{U_1U_1'+U_2U_2'}{\sqrt{U_1^2+U_2^2+h}}\\
\theta_u&= \frac{ \sqrt{(U_1 U_2'-U_2 U_1')^2+h(U_1^{\prime2}+U_2^{\prime2})}}{U_1^2+U_2^2+h}
\end{split}
\end{align}
finally:
\begin{align}\label{1.2.4}
\begin{split}
E&=(V_1+U_3)^2 ( f_u^2+f^2\theta_u^2)\\
F&=(V_1+U_3) V_1'  f f_u\\
G&= f^2 V_1^{\prime2}+V_2^{\prime2}-h V_1^{\prime2}
\end{split}
\end{align}
It is easy to check that the first fundamental form does not depend on the parameter h, which means it is an isometry (\ref{1.2.5}).
\begin{align}\label{1.2.5}
\begin{split}
E&=(V_1+U_3)^2 \Big(\frac{(U_1U_1'+U_2U_2')^2}{U_1^2+U_2^2+h}+(U_1^2+U_2^2+h)  \frac{ (U_1 U_2'-U_2 U_1')^2+h(U_1^{\prime2}+U_2^{\prime2})}{(U_1^2+U_2^2+h)^2}\Big)\\
E&=(V_1+U_3)^2 \frac{(U_1U_1'+U_2U_2')^2+(U_1 U_2'-U_2 U_1')^2+h(U_1^{\prime2}+U_2^{\prime2})}{U_1^2+U_2^2+h} \\
E&=(V_1+U_3)^2 \frac{U_1^2U_1^{\prime2}+U_2^2U_2^{\prime2}+U_1^2 U_2^{\prime2}+U_2^2 U_1^{\prime2}+h(U_1^{\prime2}+U_2^{\prime2})}{U_1^2+U_2^2+h} =(V_1+U_3)^2(U_1^{\prime2}+U_2^{\prime2})\\
F&=(V_1+U_3) V_1'  (U_1U_1'+U_2U_2')\\
G&=V_1^{\prime2}(U_1^2+U_2^2+h)+V_2^{\prime2}-h V_1^{\prime2}=V_1'^{\prime2}(U_1^2+U_2^2)+V_2^{\prime2}\\
\end{split}
\end{align}

\subsection{Pure bending energy of Goursat surface}
The second fundamental form reads:
\begin{align}
\begin{split}
L&=\frac{det(S_{uu},S_u,S_v)}{\sqrt{EG-F^2}}\label{1.3.1.a}
\end{split}
\end{align}
\begin{align}
\begin{split}
M&=\frac{det(S_{uv},S_u,S_v)}{\sqrt{EG-F^2}}\label{1.3.1.b}
\end{split}
\end{align}
\begin{align}
\begin{split}
N&=\frac{det(S_{vv},S_u,S_v)}{\sqrt{EG-F^2}}\label{1.3.1.c}
\end{split}
\end{align}
where $S_{uu}$, $S_{uv}$ and $S_{vv}$ are:
\begin{align}\label{1.3.2}
\begin{split}
S_{uu}&=V_1 ( (f_{uu} -f\theta_u^2) e_{r}+(2f_u\theta_u+f\theta_{uu}) e_{\theta})+\frac{U_3'}{V_1+U_3}S_u\\
S_{uv}&=V_1' ( f_u e_r+f\theta_u e_{\theta})\\
S_{vv}&= f V_1'' e_r+  \frac{V_2''V_2'-h V_1''V_1'}{\sqrt{V_2^{\prime2}-h V_1^{\prime2}} }e_z
\end{split}
\end{align}
with:
\begin{align}\label{1.3.3}
\begin{split}
 f_{uu}&=\frac{U_1U_1''+U_2U_2''+U_1^{\prime2}+U_2^{\prime2}}{\sqrt{U_1^2+U_2^2+h}}-\frac{(U_1U_1'+U_2U_2')^2}{\sqrt{U_1^2+U_2^2+h}^3}\\
 \theta_{uu}&= \frac{((U_1 U_2'-U_2 U_1')(U_1 U_2''-U_2 U_1'')
 +h(U_1'U_1''+U_2'U_2''))}{f^2 \sqrt{(U_1 U_2'-U_2 U_1')^2+h(U_1^{\prime2}+U_2^{\prime2})}}-\frac{\theta_u(U_1U_1'+U_2U_2')}{(U_1^2+U_2^2+h)^2} 
 \end{split}
\end{align}
 \\\\
\noindent The first determinant (\ref{1.3.1.a}) can be expanded  along the third column:
\begin{align}\label{1.3.4}
\begin{split}
det(S_{uu},S_u,S_v)&=(V_1+U_3)^2\left|\begin{array}{ccc} (f_{uu} -f\theta_u^2) &  f_u  & V_1'  f\\
(2f_u\theta_u+f\theta_{uu}) &  f\theta_u  & 0\\
0&0&   \sqrt{V_2^{\prime2}-h V_1^{\prime2}}  \end{array}\right|\\
det(S_{uu},S_u,S_v)&=(V_1+U_3)^2 \sqrt{V_2^{\prime2}-h V_1^{\prime2}}\delta
 \end{split}
\end{align}
where $\delta=(f_{uu} -f\theta_u^2)f\theta_u-f_u (2f_u\theta_u+f\theta_{uu}) $.
\\
\noindent The second determinant  (\ref{1.3.1.b}) is zero as $S_{uv}$ is collinear to $S_{u}$ (\ref{1.3.2}, \ref{1.2.2}):
\begin{align}\label{1.3.5}
\begin{split}
det(S_{uv},S_u,S_v)=0
 \end{split}
\end{align}
\\
\noindent The third determinant  (\ref{1.3.1.c}) can be expanded along the third column:
\begin{align}\label{1.3.6}
\begin{split}
det(S_{vv},S_u,S_v)&=\left|\begin{array}{ccc}f V_1''  & (V_1+U_3)f_u  & V_1'  f\\
0&  (V_1+U_3)f\theta_u  & 0\\
 \frac{V_2''V_2'-h V_1''V_1'}{\sqrt{V_2^{\prime2}-h V_1^{\prime2}} }&0&   \sqrt{V_2^{\prime2}-h V_1^{\prime2}}  \end{array}\right|\\
 det(S_{vv},S_u,S_v)&=(V_1+U_3)\Big(-V_1'  f f\theta_u \frac{V_2''V_2'-h V_1''V_1'}{\sqrt{V_2^{\prime2}-h V_1^{\prime2}} }+  \sqrt{V_2^{\prime2}-h V_1^{\prime2}}(f\theta_u)f V_1''\Big)\\
 det(S_{vv},S_u,S_v)&=f^2 \theta_u \frac{(V_1+U_3)V_2'(V_1''V_2' -V_1'V_2'')}{\sqrt{V_2^{\prime2}-h V_1^{\prime2}} }
 \end{split}
\end{align}

\noindent At the end:
\begin{align}\label{1.3.7}
\begin{split}
L&=\frac{(V_1+U_3)^2 \sqrt{V_2^{\prime2}-h V_1^{\prime2}}\delta}{\sqrt{EG-F^2}}\\
M&=0\\
N&=f^2 \theta_u \frac{(V_1+U_3)V_2'(V_1''V_2' -V_1'V_2'')}{\sqrt{V_2^{\prime2}-h V_1^{\prime2}}(\sqrt{EG-F^2}) }\\
 \end{split}
\end{align}
\\\\
The principal curvature can now be expressed:
\begin{align}\label{1.3.9}
\begin{split}
\kappa_1=H+\sqrt{H^2-K}\\
\kappa_2=H-\sqrt{H^2-K}
 \end{split}
\end{align}\\
where $H$ is the mean curvature and $K$ is the gaussian curvature:
\begin{align}\label{1.3.8}
\begin{split}
K=\frac{LN-M^2}{EG-F^2}\\
H=\frac{LG+NE}{2(EG-F^2)}
 \end{split}
\end{align}

\noindent It eventually leads to the pure bending energy:
\begin{align}\label{1.3.10}
\begin{split}
E_b=\int_u\int_v B\Big((\kappa_1-\kappa_{1,0})^2+2\nu(\kappa_1-\kappa_{1,0})(\kappa_2-\kappa_{2,0})+(\kappa_2-\kappa_{2,0})^2\Big)\sqrt{EG-F^2}du dv
 \end{split}
\end{align}
where $\kappa_{1,0}$, $\kappa_{2,0}$ stands for the curvatures at rest state $h=0$.
The expression of the square of the element of surface can be simplified:
\begin{align}\label{1.3.11}
\begin{split}
EG-F^2&=(V_1+U_3)^2\Big(V_1^{\prime2}(U_1'U_2-U_2'U_1)^2+V_2^{\prime2}(U_1^{\prime2}+U_2^{\prime2})\Big)
 \end{split}
\end{align}

\subsection{Folding Goursat surface, isometric $C^0$ deformations} 
Isometric Goursat deformation conserves two systems of planes mutually parallel one defined by a constant $u$ and the other by a constant $v$; the simple yet original idea of this article is that any of these planes can be used as a plane for a mirror-symmetry thus providing the widest-known family of analytical isometric deformations with folds. An admissible fold for an isometrical deformation is a non-moving line joining two surfaces both deforming with an isometry.\\
Mathematically the presence of a vertical fold (mirror-plane defined by $u=cst$) can be encoded by incorporating a piece-wiese constant function $\epsilon : I\to\{-1,1\}$ into the formula for $\theta$.
\begin{align}\label{1.1.2}
\begin{split}
 \forall h\geqslant0,\forall u\in I,\\
\theta^\epsilon(u,h)&=\int_0^{u} \epsilon\frac{\sqrt{(U_1 U_2'-U_2 U_1')^2+h(U_1^{\prime2}+U_2^{\prime2})}}{f^2} du
\end{split}
\end{align}
Horizontal folds are not studied herein but could be generated by abruptly changing the \text{sign} of $z'$ at a fixed $v$.\\
Let $N$ be the the number of points $(u_{fo,i})_{i\in\{1,\cdots,\ N\}}$\footnote{$\mathbf{Notations}$: The index $fo$ in $u_{fo,i}$ stands for fold.} where $\tau=\epsilon \text{sign}(U_1 U_2'-U_2 U_1')$ switches its \text{sign} on the interval $I$. $\theta_{\epsilon,u}$ can be rewritten
\begin{align}\label{1.1.2}
\begin{split}
\theta^\epsilon_{u}=\epsilon\frac{\sqrt{(U_1 U_2'-U_2 U_1')^2}}{U_1^2+U_2^2}=\tau \frac{U_1 U_2'-U_2 U_1'}{U_1^2+U_2^2}=\tau(\arctan(\frac{U_2}{U_1}))'
\end{split}
\end{align}
The integration gives for $u\in[u_{fo,i},u_{fo,i+1}]$:
\begin{align}\label{1.1.2}
\begin{split}
\theta^\epsilon(u,0)&=\sum_{j=1}^i \tau(\frac{u_{fo,j}+u_{fo,j-1}}{2})\Big[\arctan\Big(\frac{U_2}{U_1}\Big)\Big]_{u_{fo,j-1}}^{u_{fo,j}}+ \tau(u)\Big[\arctan\Big(\frac{U_2}{U_1}\Big)\Big]_{u_{fo,i}}^{u}\\
\end{split}
\end{align}\\ 
For a general $\epsilon$, the jump of slope of $\theta^\epsilon$ at each $u_{fo,i}$ will correspond to a fold on the shell for any $h\geqslant 0$.  The dihedral angle of the fold situated along the line $u=u_{fo,i}$ is fully prescribed by the mirror-symmetry (\ref{1.4.1}, \ref{1.4.2}).\\\\

\noindent Only if $\epsilon=\text{sign}(U_1 U_2'-U_2 U_1')$, there won't be a fold at rest state ($h=0$). In this latter case, the expression for $\theta^\epsilon$ is simpler:
\begin{align}
\begin{split}
\theta^\epsilon(\cdot,0)&=\arctan(\frac{U_2}{U_1})\\
(\cos(\theta^\epsilon(\cdot,0)), \sin(\theta^\epsilon(\cdot,0)))&=\frac{(U_1,U_2)}{\sqrt{U_1^2+U_2^2}}
\end{split}
\end{align}
as well as for the expression of the rest state surface ($h=0$):
\begin{align}\label{1.1.1}
\begin{split}
\forall u\in I,\forall v&\in \Big(J\cap (V_2^{\prime2}-h V_1^{\prime2})^{-1}(]0,+\infty[) \Big),\\
 S(u,v,0)&=  V_1(v)(U_1(u),U_2(u),0)+\int_0^{u} U_3(U_1',U_2',0)+ (0,0,V_2(v))
\end{split}
\end{align}
The undeformed surface $S(\cdot,\cdot,0)$ is a $C^\infty$ surface.\\

\noindent The expression for the bending energy is not affected by the mirror-symmetry as it simultaneously changes the \text{sign} of both the fundamental forms and the principal curvatures:
\begin{align}\label{1.3.7}
\begin{split}
(L^\epsilon,M^\epsilon,N^\epsilon)&=\epsilon (L,0,N)\\
(k^\epsilon_1,k^\epsilon_2)&=\epsilon (k_1,k_2)
\end{split}
\end{align}
\\\\

\subsection{Energy associated to the fold}
Mirror-symmetry does not alter the bending energy but introduces an additional energy term associated to the folds. For $i \in\{1,\cdots,N\}$, the behavior of the $i^{th}$ fold can be described using a phenomenological energy  \cite{11}:
\begin{align}\label{1.4.1}
\begin{split}
\forall v\in &\Big(J\cap (V_2^{\prime2}-h V_1^{\prime2})^{-1}(]0,+\infty[) \Big)\\
&E^\epsilon_{fo,i}=\int_{v_{min}}^{v_{max}} B\frac{\sigma}{2}\Big(\cos(\frac{\alpha_i(v,h)}{2})-\cos(\frac{\alpha_i(v,0)}{2})\Big)^2\sqrt{G(u_{fo,i},v)}dv
 \end{split}
\end{align}
where B is the bending modulus; $\sigma$ is a  constant depending on the material property as well as of the thickness along the fold; $\alpha_i(v,h)$ is the angle between the normal on both sides of the fold at the point of coordinates $(u_{fo,i},v)$. This energy has been experimentally validated at the first order for broader range of materials and dependence in thickness has been tested \cite{12}. \\
\begin{align}\label{1.4.2}
\begin{split}
E^\epsilon_{fo,i}=\int_{v_{min}}^{v_{max}}  B\frac{\sigma}{2} \Big( \sqrt{\frac{\cos(\alpha_i(v,h))+1}{2}}- \sqrt{\frac{\cos(\alpha_i(v,0))+1}{2}}\Big)^2\sqrt{G(u_{fo,i},v)}dv
 \end{split}
\end{align}
In the case where $\epsilon=\text{sign}(U_1'U_2-U_2'U_1)$, $\alpha_i(v,0)=0$ as there is no fold for $h=0$.
 
\noindent For a given $v$, the dot product between $\mathcal{N}^-$ the normal to the surface calculated  in $(u_{fo,i}^-,v)$ on one side of the fold and $\mathcal{N}^+$ the normal in $(u_{fo,i}^+,v)$ on the other side gives $\cos(\alpha_i(v,h))$:
\begin{align}\label{1.4.3}
\begin{split}
\mathcal{N^+}&=\frac{S^+_u \times S_v}{\sqrt{EG-F^2}}\\
\mathcal{N^+}&=\frac{(V_1+U_3) ( f_u e_r-f\theta_u^{\epsilon,+} e_{\theta})\times ( f V_1' e_r+  \sqrt{V_2^{\prime2}-h V_1^{\prime2}} e_z)}{\sqrt{EG-F^2}}\\
\mathcal{N^+}\cdot \mathcal{N^-}&=(V_1+U_3)^2\frac{ (f^2\theta_u^{\epsilon,+}\theta_u^{\epsilon,-} (V_2^{\prime2}-h V_1^{\prime2} )+ f_u^2 (V_2^{\prime2}-h V_1^{\prime2}) +f^4\theta_u^{\epsilon,+} \theta_u^{\epsilon,-} V_1^{\prime2} )}{EG-F^2}\\
\cos(\alpha_i(v,h))&=(V_1+U_3)^2\frac{ (f^2\theta_u^{\epsilon,+}\theta_u^{\epsilon,-}  (V_2^{\prime2}-h V_1^{\prime2} )+ f_u^2 (V_2^{\prime2}-h V_1^{\prime2}) +f^4\theta_u^{\epsilon,+} \theta_u^{\epsilon,-}  V_1^{\prime2} )}{EG-F^2}
 \end{split}
\end{align}

\noindent The total energy reads:
\begin{align}\label{1.4.4}
\begin{split}
E_{tot}&=\Sigma _{i=1}^{i=N}E^\epsilon_{fo,i}+E_b
 \end{split}
\end{align}

\section{Isometric deformations of a banana-shaped family of surfaces}
\subsection{Geometrical family of banana-shaped surfaces}
\begin{figure}[h!]
\begin{center}
\includegraphics[width=5in]{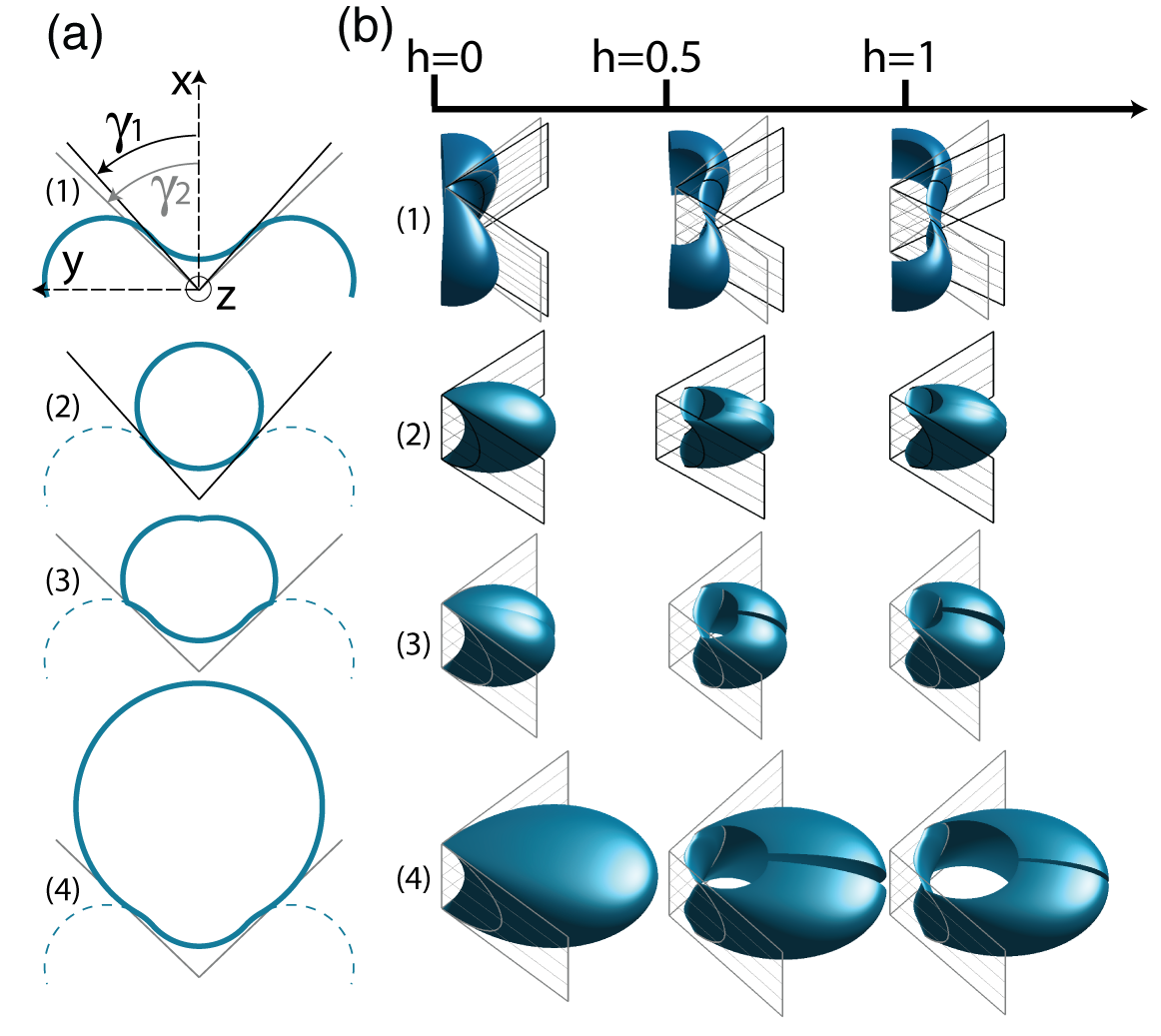}
\caption{(a) Each panel (1-4) represents the $z=0$-trace, or intersection between the horizontal plane $z=0$ and a surface of the family at rest state ($h=0$). Some of the parameters are shared by all the surfaces: $R_1=1$, $\alpha_1=  0.7752$ and $b=1$. In each case, we choose $u_{\max}=u_{cl,0}$ to close the contour. The trace of the Goursat surface ($\epsilon=1$) is represented by a full-blue line on (1) and by a dashed-blue line on (2)-(3)-(4). The black (resp. gray) straight line forming an angle $\gamma_1$ (resp. $\gamma_2$) with the  $x$-axis correspond to the trace of the black (resp. gray) mirror-plane  in the plane $z=0$: \newline
(1) $\alpha_1=\alpha_2$ and $\epsilon=1$, it is a Goursat surface without fold; \newline
(2) $\alpha_1=\alpha_2$, $\gamma_1=\theta^{\epsilon}(u_{fo,1},0)=\theta^{\epsilon}(u_{jun,1},0)$, there is no fold as the the mirror-plane is tangent to the surface; \newline
(3) $\alpha_1=\alpha_2$,  $\gamma_2=\theta^{\epsilon}(u_{fo,1},0)>\theta^{\epsilon}(u_{jun,1},0)$, there is two folds (located at the intersection between the gray straight lines and the blue contour) as the mirror-plane is not tangent to the surface; \newline
(4) $\alpha_1>\alpha_2$, $\gamma_2=\theta^{\epsilon}(u_{fo,1},0)=\theta^{\epsilon}(u_{jun,2},0)$, there is no fold as the mirror-plane is tangent to the surface. \newline
(b) Family of isometric deformations corresponding to the traces in (a) with $c=10$ and $d=10$. The deformation increases from left to right ($h=0$, $0.5$, $1$). The black (resp. gray) plane corresponds to the mirror-plane containing the $z$-axis and making an angle $\gamma_1$ (resp. $\gamma_2$) with the $x$-axis. The curves in black (resp. gray) are the intersections of the banana-shaped surface with the mirror-planes.\newline
(1) The surface opens with increasing $h$. \newline
(2) The presence of the fold counteracts the opening at the saddle point. The surface closes on itself while increasing its longitudinal deformation.\newline
(3) As the fold is further away along the surface, the opening dominates the closing effect of the fold.\newline
(4) As the fold is further away along the surface, the opening dominates the closing effect of the fold.}
\label{Figure_2}
\end{center}
\end{figure}

Let us choose the following parameters: $R_1>0$, $\alpha_1\in\ [0,\frac{\pi}{2}]$, $\alpha_2\in\ [0,\alpha_1],\alpha_2\neq\frac{\pi}{2}$, $b>0$, $c>0$, $d>0$. We introduce the two circles $\mathcal{C}_1$,  $\mathcal{C}_{1,r}$  of identical radius $R_1$ centered in 
\begin{align}
\begin{split}
(x_{1},\ y_1)&=R_1\Big(\frac{1}{\cos(\alpha_1)},\ 0\Big)\\
(x_{1,r},\ y_{1,r})&=R_1\Big(\frac{1}{\cos(\alpha_1)}-2\cos(\alpha_1),\ 2\sin(\alpha_1)\Big)
 \end{split}
\end{align}
and the circle $\mathcal{C}_2$ of radius  $R_2=(\frac{2\sin(\alpha_1)}{\sin(\alpha_2)}-1)R_1$  centered in
\begin{align}
\begin{split}
(x_{2},y_{2})=(x_{1,r}+(R_1+R_2)\cos(\alpha_2),\ 0).
\end{split}
\end{align}
\\ By construction, $\mathcal{C}_1$ is tangent to $\mathcal{C}_{1,r}$, and  $\mathcal{C}_{1,r}$  is tangent to $\mathcal{C}_2$; a $C^1$ curve $(U_1,U_2)$ can be constituted by joining arcs of these three circles at the two tangent intersections (Figure \ref{Figure_2}a). We note: $u_{jun,1}\footnote{The index $jun$ stands for "junction".}=R_1\alpha_1,\ u_{jun,2}=u_{jun,1}+R_1(\alpha_1-\alpha_2), \ u_{cl,0}\footnote{The index $cl$ stands for "close": $u_{cl,h}$ closes the contour at deformation $h$} =u_{jun,2}+R_2(\pi-\alpha_2)$.  $U_1$ and $U_2$ are piecewise functions on $[0,u_{cl,0}]$ defined by:
\begin{align}\label{2.1}
\begin{split}
\forall u\in&[0,u_{jun,1}],\\
&(U_1(u),U_2(u))=\Big(x_1+R_1\cos(\pi-\frac{u}{R_1}),\ bR_1\sin(\pi-\frac{u}{R_1})\Big),\\
\forall u\in&[u_{jun,1},u_{jun,2}],\\
&(U_1(u),U_2(u))=\Big(x_{1,r}+R_1\cos(\frac{u-u_{jun,1}}{R_1}),\ b(y_{1,r}+R_1\sin(\frac{u-u_{jun,1}}{R_1}))\Big),\\
\forall u\in&[u_{jun,2},u_{cl,0}],\\
 &(U_1(u),U_2(u))=\Big(x_2+R_2\cos(\pi-\alpha_2-\frac{u-u_{jun,2}}{R_2}),\ bR_2\sin(\pi-\alpha_2-\frac{u-u_{jun,2}}{R_2})\Big)
\end{split}
\end{align}
\\ 
$b$ is an additional parameter tuning the ellipticity of the shape. In the case $\alpha_2=\alpha_1$,\ the curve $(U_1,U_2)$ coincides with the circle $\mathcal{C}_1$.\\
The additional function $U_3$ is chosen null on $[0,Êu_{cl,0}]$, and $V_1$, $V_2$ are defined by:
\begin{align}\label{2.1}
\begin{split}
\forall v\in &[-\frac{\pi}{2},\frac{\pi}{2}],\\
Ê&(V_1(v),V_2(v))=(c\cos(v), d\sin(v))
\end{split}
\end{align}
If either $\epsilon=\text{sign}(u_{jun,1}-u)$ or $\epsilon=\text{sign}(u_{jun,2}-u)$, the folded Goursat surface has no apparent fold at the undeformed state; if additionally $\alpha_2=\alpha_1$, the rest state surface is $C^\infty$. \\

\noindent The behavior of the surfaces depends on the presence of the folds and their location: the Goursat isometric family of deformations ($\epsilon=1$) opens while increasing the longitudinal curvature at the saddle point (Figure \ref{Figure_2}b (1)); the presence of the fold tends to counteract this effect and to close the shell  (Figure \ref{Figure_2}b (2)),  but provided the fold is far enough from the junction point ($u_{fo,1}$ sufficiently bigger than $u_{jun,1}$) the opening dominates the closure for small enough deformations and biomimetic thin shells can be devised (Figure \ref{Figure_2}b (3-4)). The upper and lower extremities of the surface progressively become imaginary when $h$ increases and thus progressively disappear on Figure \ref{Figure_2}b.

\subsection{De\text{sign} of a $C^{\infty}$ thin shell transferring a vertical closing movement into an horizontal closing movement}
\begin{figure}[h!]
\begin{center}
\includegraphics[width=5in]{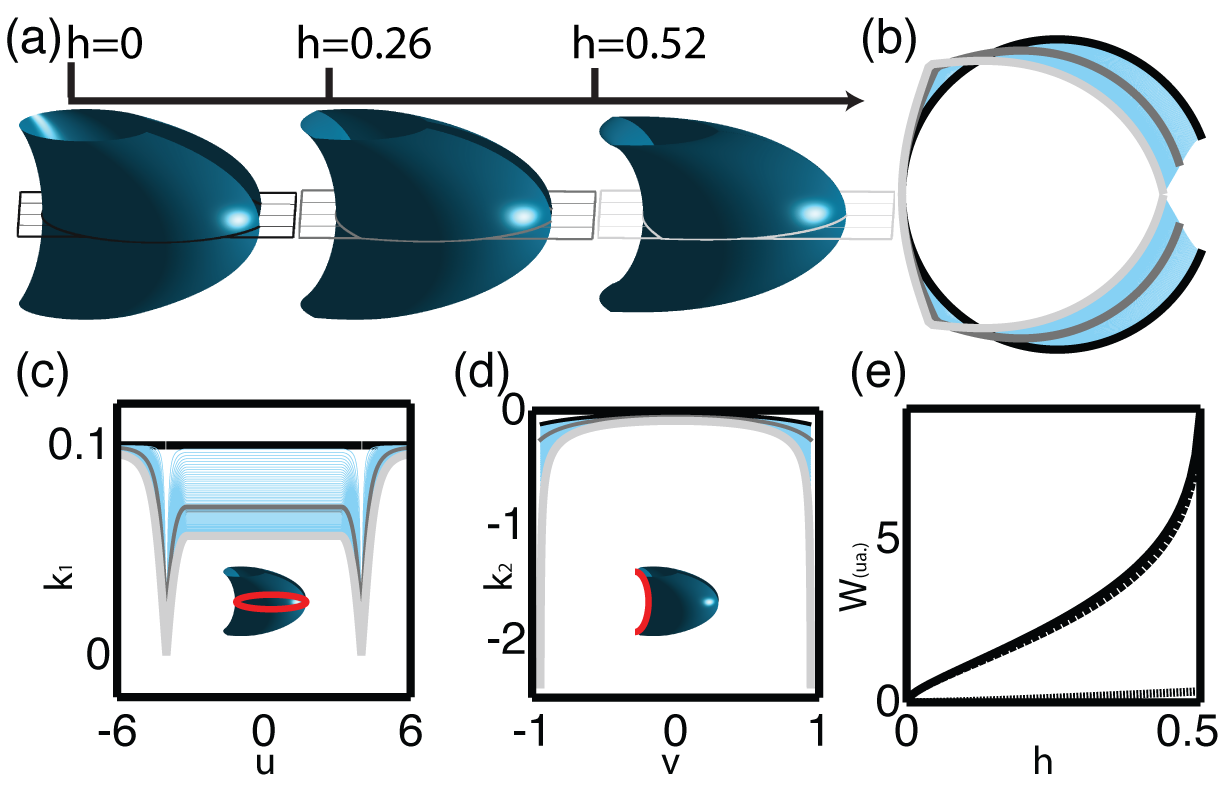}
\caption{(a) Isometric deformations of a banana-shaped surface ($R_1=1$,$\alpha_1=\alpha_2= 0.7752$, $b=1$, $c=10$, $d=10$, $v_0=0.94$, $\epsilon=\text{sign}(u_{jun,1}-u)$, $h\in\{0,\ 0.26,\ h_{max}=0.52\}$). The width of the aperture has been calculated to close the shell at the maximal deformation corresponding to the initial surface. (b) Horizontal section of the surface: the family of isometries behaves similarly at the saddle point as the seedpod but not globally. Due to the fold, the decrease in meridional curvature triggers a closing rather than an opening of the shell. The red, green and yellow contours correspond to the intersecting planes in the (a) panel. (c) Meridional curvature in the horizontal plane of symmetry: the meridional curvature decreases in absolute value with the deformations.  (d)   Longitudinal curvature in the vertical plane of symmetry: the longitudinal curvature increases in absolute value with the deformations. It diverges at the lower and upper boundary when reaching the maximal deformation. (e) Mechanical energy: Pure bending energy (dotted line). Fold energy  (dashed line). Total energy (Plain line). The bending energy diverges at $h_{max}$ while the fold energy remains bounded. In this particular example, $0=\cos(\theta(u_{cl,h_{max}},h_{max}))$ and  $\sigma= \frac{2}{\sqrt{3}}$ as in \cite{10}.}
\label{Figure_3}
\end{center}
\end{figure}

The family of isometries derived herein is an interesting tool to de\text{sign} self-sealing thin shells. For instance, given  $v_0\in[0,\frac{\pi}{2}]$ the size of the aperture $u_{c}$ which seals the shell at a prescribed vertical deformation $$\lambda=\frac{z_{c}}{z_0}$$ can be worked out  by simply looking for the solutions $(u_{cl,h_{\lambda}},h_{\lambda})$ of  (\ref{2.2}) (Figure \ref{Figure_3}(a)(b)).
\begin{align}\label{2.2}
\begin{split}
0&=\theta(u_{cl,h_{\lambda}},h_{\lambda})\\
\lambda&=\frac{ \int_0^{v_{0}} \sqrt{d^2\cos(v)^2-h_{\lambda}c^2 \sin(v)^2} dv}{c\sin(v_{0})}
\end{split}
\end{align}
\\
\noindent The quantitative behavior of the curvatures at the saddle point can also  be observed thanks to (\ref{1.3.9}): the meridional curvature decreases in absolute value  (Figure \ref{Figure_3}(c))  while the longitudinal curvature increases. When reaching $$h_{max}=\frac{d^2}{c^2\tan(v_{0}^2)},$$ the longitudinal curvature $k_2$ diverges (Figure \ref{Figure_3}(d)) and the bending energy (\ref{1.3.10}) as well (Figure \ref{Figure_3}(e)). On the contrary the energy associated to the fold (\ref{1.4.3}) remains bounded throughout the deformation.

\subsection{De\text{sign} optimization of the $C^1$ biomimetic thin shells family transferring a vertical closing movement into an horizontal opening movement}
\begin{figure}[h!]
\begin{center}
\includegraphics[width=5in]{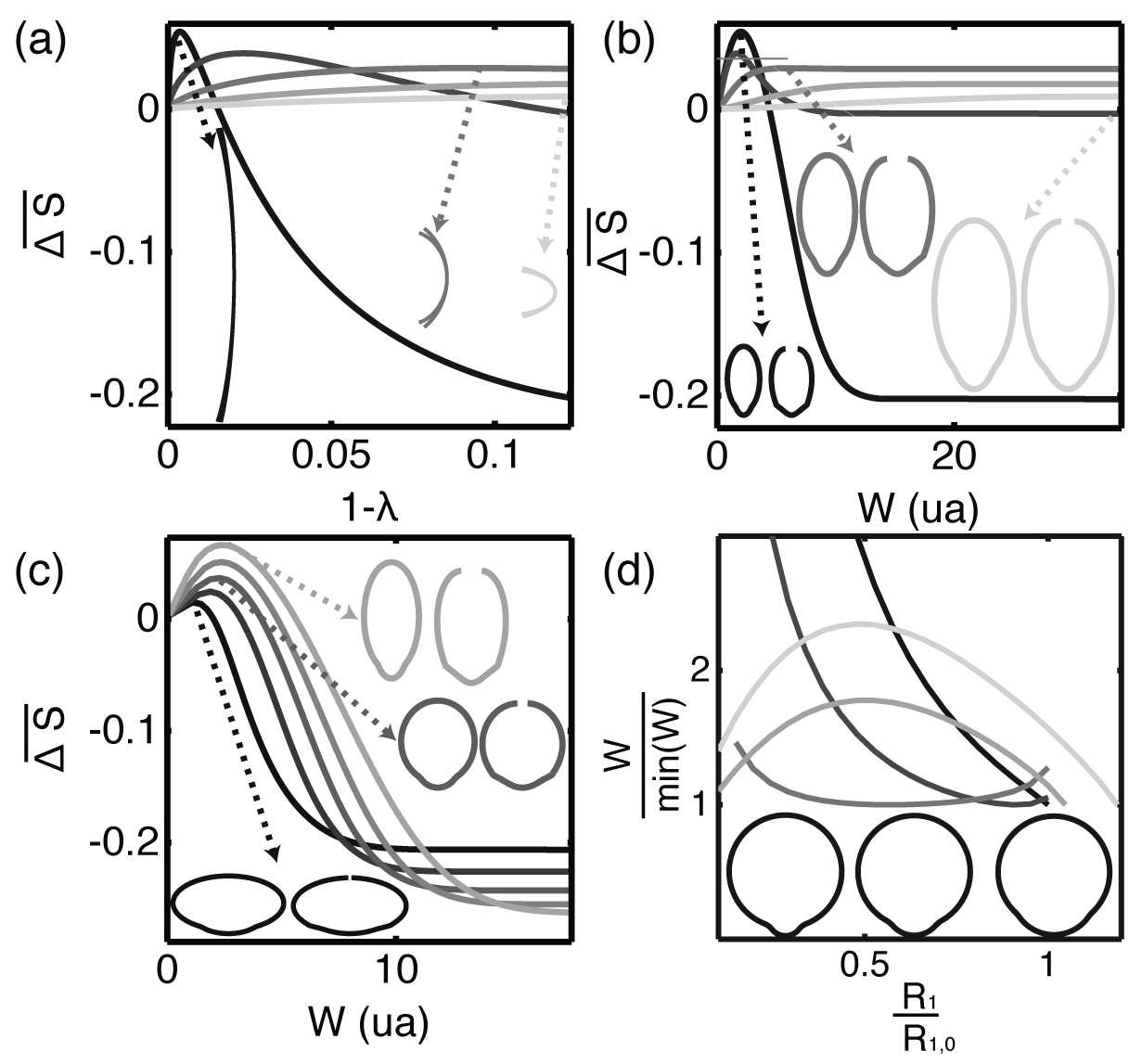}
\caption{(a,b) Influence of $x-z$ shape, the rest state volume $Vol$ being identical. $R_1\in\{0.71,0.96,1.23,1.48,1.73\}$, $\alpha_1=0.54$, $\alpha_2=0.37$, $b=0.5$, $\ c=10$, $d=\frac{10}{R_1^2}$  (the darkness of the gray of the line in the plot increases with $d$) $, v_0= 0.94, 0<h<\frac{d^2}{c^2\tan(v_{0}^2)}$.  \newline
(a) Normalized opening  $\overline{\Delta S}$ versus vertical deformation $1-\lambda$.  The arrows point to the corresponding rest state contour $(x(0,.,0),z(0,.,0))$ and deformed contour  $(x(0,.,h),z(0,.,h))$ at maximal opening.(b) 
Normalized opening  $\overline{\Delta S}$ versus energetical cost $W$. The arrows point to the corresponding rest state contour $(x(0,.,0),y(0,.,0))$ and deformed contour  $(x(0,.,h),y(0,.,h))$ at maximal opening (The scale is 0.3 smaller than for the $x-z$ curve on (a)). (c,d) Influence of $x-y$ shape: (c) At constant $Vol$, normalized opening  $\overline{\Delta S}$ versus energetic cost $W$. $R_1\in\{0.71,0.96,1.23,1.48,1.73\}$ , $\alpha_1=0.54$, $\alpha_2=0.37$, $b=\frac{1}{R_1^2}$  (the darkness of the gray of the line in the plot increases with $b$)$,\ c=10,\ d=50, v_0= 0.94, 0<h<\frac{d^2}{c^2\tan(v_{0}^2)}$. The arrows point to the corresponding rest state contour $(x(0,.,0),y(0,.,0))$ and deformed contour  $(x(0,.,h),y(0,.,h))$.\newline
(d) Energetic cost normalized by its minimum versus $R_1$ for constant $\overline{\Delta S}$ and constant rest state volume $Vol$.  $R_{1,0}\in \{0.71,0.96,1.23,1.48,1.73\} $, $0.1<\frac{R_{1}}{R_{1,0}}<1.9$ ,$\alpha_{1,0}=0.54$, $\alpha_{2,0}=0.37$, $b=1$, $c=10$, $d=\frac{10}{R_{1,0}^2}$, $v_0= 0.94$, $0<h<\frac{d^2}{c^2\tan(v_{0}^2)}$. The contours at the bottom are the rest state contours $(x(0,.,0),y(0,.,0))$ while $R_1$ is varying.}
\label{Figure_4}
\end{center}
\end{figure}

Constraint optimization can be carried on the parameters with cost function either the energy or the vertical deformation and utility function the opening area $\Delta S$ defined by:
\begin{align}
\begin{split}
\Delta S(R_1,\alpha_1,\alpha_2, b, c, d,v_0,h)=\int_{v=-v_0}^{v=+v_0}y(u_{cl,0},v,h)z'(v)dv
\end{split}
\end{align}
As the rest state surface is closed, the inner volume $Vol$ is well-defined. Changing $R_1,\ b,\ c,\ d$  corresponds to applying an affine transformation on the rest state surface which modifies $Vol$ according to:
\begin{align}
\begin{split}
Vol(R_1,\alpha_1,\alpha_2, b, c, d,v_0)=R_1^2 b c^2 d\ Vol(1,\alpha_1,\alpha_2, 1, 1, 1,v_0)
\end{split}
\end{align}
For this reason, the constant $Vol$ is a convenient hard constraint in the minimization: modifying a parameter can easily be compensated by inversely changing another one. For the constant $Vol$ minimization, it is convenient to normalize the opening: $\overline{\Delta S}=\frac{S}{Vol^{\frac{2}{3}}}$.\\
Another possible hard constraint for the minimization is $\lambda$ the vertical deformation defined by:
\begin{align}
\begin{split}
\lambda(R_1,\alpha_1,\alpha_2, b, c, d,v_{v_0},h)=\frac{z(v_0,h)}{z(v_0,0)}\\
\end{split}
\end{align}
For $h_{max}=\frac{d^2}{c^2\tan(v_{0}^2)}$, $\lambda_{max}$ the maximal deformation after which the upper part of the surface becomes imaginary simplifies into:
\begin{align}
\begin{split}
\lambda(R_1,\alpha_1,\alpha_2, b, c, d,v_0,h_{max})&=\frac{\int_{v=0}^{v=v_0} \sqrt{d^2\cos(v)^2-\frac{d^2}{\tan(v_0)^2c^2}c^2\sin(v)^2}dv}{d\sin(v_0)}\\
\lambda(R_1,\alpha_1,\alpha_2, b, c, d,v_0,h_{max})&=\frac{\int_{v=0}^{v=v_0} \sqrt{\cos(v)^2-\frac{\sin(v)^2}{\tan(v_0)^2}}dv}{\sin(v_0)}
\end{split}
\end{align}
$\lambda_{max}$ is thus independent of both $d$ and $c$ but depends only of $v_0$.\\\\
\noindent The influence of some of the parameters have been studied around an initial shape determined by the parameter ($R_{1,0}$, $\alpha_{1,0}$, $\alpha_{2,0}$, $b_0$, $c_0$, $d_0$, $v_0$):\\
- The influence of the $y=0$ trace (or $x-z$ contour) on opening efficiency has been assessed. The parameter $d$ was made to vary while maintaining $Vol$ constant by adjusting $R_1 =R_{1,0} \sqrt{\frac{d_0}{d}}$. If increasing $d$ does not affect $\lambda_{max}$, it increases $h_{\max}$ and thus the maximal $x-y$ deformation. Provided $d$ is sufficient, the movement first consists in an opening dominated by the saddle point ($\overline{\Delta S}>0$, Figure \ref{Figure_4}a), and then in a closure dominated by the fold ($\overline{\Delta S}<0$, Figure \ref{Figure_4}a); for smaller $d$, the shell only opens as the top of the shell becomes imaginary before reaching the closing phase. Increasing $d$ increases the opening amplitude (Maximum of $\overline{\Delta S}$ on Figure \ref{Figure_4}a) and decreases the vertical deformation necessary to trigger this maximum; increasing $d$ also decreases the energetic cost of the maximal opening (Figure \ref{Figure_4}b). The elongated shapes are the most efficient to trigger an opening according to both criteria of vertical deformation efficiency and energetic cost.\\
- The influence of the $z=0$ trace  (or $x-y$ contour)  on opening efficiency has been assessed: We modified the ellipticity $b$ while maintaining $Vol$ constant by adjusting $R_1 =R_{1,0} \sqrt{\frac{b_0}{b}}$. Decreasing $b$ increases the amplitude of the maximal opening and decreases the energetic cost of it (Figure \ref{Figure_4}c). The most efficient shapes are obtained for ellipsis elongated in the $x$ direction ($b$ small). A second minimization was carried out with two hard constraints: constant opening and constant initial volume. The influence of the even repartition of the curvature  along the $x-y$ closed profile on the energetic cost was investigated. Practically  $R_1$ was modified while both $Vol$ and $\Delta S$ were maintained constant by only adjusting  $\alpha_1$, $\alpha_2$. For each $R_1$, $\alpha_1$ and $\alpha_2$ were obtained by solving the two following equations with fsolve of Matlab:
\begin{align}
\begin{split}
Vol(1,\alpha_1,\alpha_2, 1, 1, 1,v_0)&=\Big(\frac{R_{1,0}}{R_1}\Big)^2Vol(1,\alpha_{1,0},\alpha_{2,0}, 1, 1, d,v_0)\\
\Delta S(R_1,\alpha_1,\alpha_2, b, c, d,v_0,h)&=\Delta S(R_{1,0},\alpha_{1,0},\alpha_{2,0}, b, c, d,v_0,h)
\end{split}
\end{align}
The location of the minimum for the energetic cost strongly depends on the $x-z$ shape (Figure \ref{Figure_4}d): for $d>>1$ high enough, the optimum is located at the highest $R_1$ which corresponds to most homogeneous repartition of curvature (see the $z=0$ trace rest state profile at the bottom of the Figure  \ref{Figure_4}d);  for  $d\approx1$, the minimum shifts  to lower $R_1$ which means more even distribution of curvature; for smaller $d$ there are two optima, one with the least homogeneous distribution of curvature ($R_1$ minimal) and one with the most homogeneous distribution of curvature ($R_1$ maximal). 

\section{Conclusion}
Isometric families of deformation were traditionally one of the few ways to provide analytical examples for the deformation of thin shells. For this reason the geometers and mechanicians of the late nineteenth century and early  twentieth century have intensely looked for such solutions \cite{9},\cite{13}. Goursat surfaces were the onset of 40 years of research on isometry; a few years later Tzizeica \cite{14} discovered his own independent family of surfaces; Drach \cite{15} and then Gambier \cite{16} finally noticed that all the known families of isometric deformations were particular solutions of some Moutard equations verifying additional conditions. While the subsequent research improved the integration of Moutard equations which are now known to be fully analytical, it did not lead to any new simple family of analytical isometric deformations \cite{17}. Some of these results were proposed by Eisenhart as exercises without the references of the authors \cite{18}. This work is nowadays quite forgotten and not even mentioned in modern textbooks. In this context, this article provides a new one-parameter family of $C^0$ surfaces extending the classical $C^\infty$ family of  Goursat by naturally adding curved folds. A  closed form for the energy associated to its deformation is for the first time provided. This theoretical  family of surfaces is illustrated by two examples of banana-shaped surfaces: a shell transforming a vertical closure into an horizontal sealing, and a biomimetic shell triggering an horizontal opening by a vertical closure. In order to actually trigger these modes of deformation in experiments, the fold area has probably to be elastically softer than the remaining part of the shell.\\
Research for smart materials and actuators has prompted the need for better theoretical tools to describe the interplay between shells and curved folds. Important progresses have recently been obtained for growing \cite{19} and inert thin shells \cite{11},\cite{20} with both straight and curved folds. Nevertheless numerical codes are not yet fully validated. In this context even if analytical solutions can describe only restricted modes of deformation, they are of great use to check the accuracy of simulations especially in such exotic examples.\\
Vegetal thin shells e.g. pollen grains and seedpods are a biomimetic source of inspiration for packaging designers.  Analytical families of deformations are obtained much more rapidly than simulations; as such they constitute a valuable guide for designers enabling them to explore the potentialities of different shapes and to design precisely the apertures. As illustrated herein, the easiness or the difficulty for closing the different shells can be very easily assessed thanks to the pure bending energy; the actual energy necessary to trigger a given deformation is slightly lowered by some stretching which nevertheless corresponds to very mild geometrical distortions \cite{5}, \cite{6}.\\

\noindent $\mathbf{Acknowledgement.}$ I thank Jacques Dumais and Enrique Cerda for providing me the example of the seedpod of $\textit{Accacia caven}$.\\

\noindent $\mathbf{Funding.}$ Etienne Couturier has been funded by CNRS and by the Fondecyt postdoctoral fellowship 3120105.


\end{document}